\documentclass[pdflatex,sn-mathphys-num]{sn-jnl}

\usepackage{graphicx}%
\usepackage{multirow}%
\usepackage{amsmath,amssymb,amsfonts}%
\usepackage{amsthm}%
\usepackage{mathrsfs}%
\usepackage[title]{appendix}%
\usepackage{xcolor}%
\usepackage{textcomp}%
\usepackage{manyfoot}%
\usepackage{booktabs}%
\usepackage{algorithm}%
\usepackage{algorithmicx}%
\usepackage{algpseudocode}%
\usepackage{listings}%

\theoremstyle{thmstyleone}%
\newtheorem{theorem}{Theorem}
\theoremstyle{thmstyletwo}%

\theoremstyle{thmstylethree}%

\def\Phat{\hat{P}}
\def\Mhat{\hat{M}}
\def\Hhat{\hat{H}}
\def\Qhat{\hat{Q}}
\def\deltahat{\hat{\delta}}
\def\epsilonhat{\hat{\epsilon}}
\def\gammahat{\hat{\gamma}}
\def\muhat{\hat{\mu}}
\def\lambdahat{\hat{\lambda}}
\def\K{\mathcal{K}}

\def\E{\mathrm E}
\def\argmin{\mathop\mathrm {argmin}}
\def\tr{\mathop\mathrm {tr}}

\def\Var{\mathop\mathrm {Var}}
\def\Mean{\mathop\mathrm {Mean}}

\begin{document}

\title[Article Title]{Cartesian Statistics on Spheres}

\author*{\fnm{Rudolf} \sur{Beran}} \email{rjberan@ucdavis.edu}

\affil{\orgdiv{Department of Statistics}, \orgname{University of California, Davis}, \country{USA}}

\abstract{Directional data consists of unit vectors in \textit{q}-dimensions that can be described in polar or Cartesian coordinates. Axial data can be viewed as a pair of directions pointed in opposite directions or as a projection matrix of rank 1. Historically, their statistical analysis has largely been based on a few low-order exponential family models of distributions for random directions or axes. A lack of tractable algebraic forms for the normalizing constants has hindered the use of higher-order exponential families for less constrained modeling. Of interest are functionals of the unknown distribution of the directional/axial data, such as the directional/axial mean, dispersion, or distribution itself.  This paper outlines nonparametric estimators and bootstrap confidence sets for such functionals. The procedures are based on the empirical distribution of the directional/axial sample expressed in Cartesian coordinates. Sketched as well are nonparametric comparisons among multiple mean directions or axes, estimation of trend in mean directions, and analysis of \textit{q}-dimensional observations restricted to lie in a specified compact subset.}


\maketitle 

\section{Introduction} \label{sec1} 

Geological data often includes directions or axes measured in three
dimensions. Examples are directions of remanent magnetization of lava
cores, axes normal to geological folding planes, or positions on the surface
of the earth such as that of the north or south magnetic poles. Biological
measurements can consist of directions in two dimensions. Instances are
directions in which birds or bees fly after release or the time-of-day viewed
as a circular variable. In econometrics, season or month are discrete circular covariates that correspond to angular parts of the earth's orbit around the sun. Statistical analysis of directional and axial data has had a rich century of development.  Specific examples of directional and axial data, of how to plot these, and of their classical parametric analyses are presented in the expository books \cite{bib13}, \cite{bib18} and \cite{bib19}. Some of the nonparametric analyses described in what follows are exemplified in the research papers \cite{bib5}, \cite{bib14}, \cite{bib16} and \cite{bib4}.

A \textit{direction} in $q$-dimensions is a $q \times 1$ unit vector 
$d$. In Cartesian coordinates it can be visualized as a point on the unit sphere $S_q = \{u \in R^q: |u|=1\}$, where $|\cdot|$ is the Euclidean norm. The equivalent expression of $d$ in polar coordinates is useful to record and plot $d$ in $2$ or $3$ dimensions. The Cartesian form supports formulation of the statistical procedures outlined in this paper.

An \textit{axis} in $q$-dimensions is an unordered pair of directions $\{e, -e\}$, where $e \in S_q$. We write $\pm e$ to represent the pair and, by convention, require the first nonzero component of $e$ be positive. An axis may be visualized as a pair of diametrically opposed points on the sphere $S_q$ or as one of those points that lies on a pre-selected hemisphere. Equivalently the axis $\pm e$ may be represented as the rank $1$ projection matrix $ee'$. Given the value of $ee'$, we recover $\pm e$ as the sign ambiguous eigenvector of $ee'$ associated with the eigenvalue $1$. The other eigenvalues are zero. The set of all axes is thus identified with the set of all orthogonal projections of $R^q$ into one-dimensional subspaces.

The mean and variance of a univariate random variable $z$ with finite second moment are defined through
\begin{equation}
\Mean (z) = \argmin_{c \in R}\E|z - c|^2 = \E(z)
\label{1.01}
\end{equation}
and
\begin{equation}
\Var(z) = \min_{c \in R}\E|z - c|^2 = \E|z - \E(z)|^2,
\label{1.02}
\end{equation}
where $\E$ is the expectation operator. These extremal characterizations suggest counterparts for directional or axial data.

To model noisy measurements of directions, let $x$ be a random unit vector with distribution $P$ on $S_q$. The mean of $P$ (or $x$) is the vector
\begin{equation}
m(P) = \E (x).
\label{1.1}
\end{equation}
Evidently $0 \le |m(P)| \le 1$. By analogy with \eqref{1.01}, the  \textit{mean direction} of $P$ is defined as
\begin{equation}   
\mu(P) = \argmin_{|\mu| = 1}\E|x - \mu|^2 = \argmin_{|\mu| = 1}(2-2\mu' m(P)) =  m(P)/|m(P)|.
\label{1.2}
\end{equation}
The last equality assumes $|m(P)| > 0$, which excludes certain highly symmetric distributions such as the uniform distribution on $S_q$.  When $|m(P)| = 0$, $\mu(P)$ is any unit vector in $S_q$.  For instance, when $P$ is the uniform distribution on $S_q$, there is no preferred mean direction.

Correspondingly, by analogy with \eqref{1.02}, the \textit{directional dispersion} of $P$ is defined as
\begin{equation}
\delta(P) = \min_{|\mu| = 1}\E|x - \mu|^2 = \E|x-\mu(P)|^2 = 
2(1-|m(P)|).
\label{1.3}
\end{equation}
This dispersion definition assesses geometrically the spread of the distribution $P$.
Evidently $0\le \delta(P) \le 2$ and $\delta(P)$ is monotone decreasing in 
$|m(P)|$. Values of $\delta(P)$ near $0$ or near $2$ indicate, respectively, low or high dispersion. The uniform distribution on $S_q$ has the maximum dispersion. In practice $\delta(P)$ may be replaced with $(1/2)\delta(P)$. 

The Euclidean norm $||\cdot||$ of a matrix $B = \{b_{ij} \colon 1 \le i,j \le  q\}$ is defined by $||B||^2 = \tr(B'B)$. A \textit{random axis} $\pm x$ is an unordered pair of random directions $x$. It can equivalently be represented by the rank $1$ projection matrix $xx'$, metrized by the norm $||\cdot||$. The form $\pm x$ is the sign ambiguous eigenvector of $x'x$ associated with the eigenvalue $1$. Let $Q$ denote the distribution of $xx'$. The mean of $Q$ is the $q \times q$ matrix
\begin{equation}
M(Q)=\E(xx').
\label{1.4}
\end{equation}
By analogy with \eqref{1.01}, the \textit{mean axis} of $Q$ is defined as $\pm \epsilon(Q)$ or 
$\epsilon(Q)\epsilon(Q)'$ with
\begin{equation}
\epsilon(Q)= \argmin_{|\epsilon| = 1}\E||xx' - \epsilon\epsilon'||^2 
=\argmin_{|\epsilon| = 1}(2-2\epsilon'M(Q)\epsilon).
\label{1.5}
\end{equation}
Thus $\pm e(Q)$ is the sign ambiguous eigenvector associated 
with the largest eigenvalue $\lambda_{\max}(Q)$ of $M(Q)$.
By analogy with \eqref{1.02}, the \textit{axial dispersion} of $Q$ is defined to be
\begin{equation}
\gamma(Q) = \min_{|\epsilon|=1}\E||xx' - \epsilon\epsilon'||^2
= 2(1 - \lambda_{\max}(Q)).
\label{1.6}
\end{equation}
In practice $\gamma(Q)$ may be replaced by $(1/2)\gamma(Q)$. 

This paper outlines nonparametric methods for estimating a directional distribution $P$ or an axial distribution $Q$ or certain functionals defined on these distributions. Section 2.1 constructs nonparametric confidence cones for mean direction, confidence intervals for directional dispersion, and confidence sets for a directional distribution.  The last are defined through the half-space metric \cite{bib12} for distributions on $R^q$. Critical values for the confidence sets are constructed with bootstrap techniques and are justified by asymptotic bootstrap theory. Section 2.2 treats analogously nonparametric confidence double cones for mean axis, confidence intervals for axial dispersion, and confidence sets for an axial distribution. Section 3 touches on extensions to extrinsic estimation of trend in mean directions, to techniques for comparing multiple mean directions, and to analysis of data restricted to lie in a specified compact subset of $R^q$.

The development of statistical methods for directional or axial data in dimensions $2$ or $3$ has historically relied on modeling the distribution $P$ or $Q$ with a {\sl parametric family} indexed by relatively few parameters. Particularly important models on $S_3$ have been the Fisher distribution, the Bingham distribution, the Kent distribution with 5 parameters, and the Fisher-Bingham distribution with $8$ parameters. The first two and the fourth of these constitute canonical exponential families on $S_3$. They all belong to a hierarchy of canonical exponential families for directional or axial distributions on $S_q$ described in \cite{bib1}. The Kent distribution is a curved submodel of the full Fisher-Bingham model. Other parametric and semiparametric directional models are treated in \cite{bib13} and \cite{bib8}. The survey paper \cite{bib15} includes engaging statistical history and anecdotes that surround emergence of the Fisher distribution.

Relations among the foregoing exponential family models are described in \cite{bib6} and \cite{bib7}. Although canonical exponential families have a strictly concave log-likelihood function and a well-articulated asymptotic theory, this advantage is offset in all but a few special cases by lack of a tractable expression for the normalizing factor of the density. The paper \cite{bib1} proposed an estimator that is asymptotically equivalent under the exponential family model to the maximum likelihood estimator without requiring knowledge of the normalizing factor. The thesis \cite{bib7} explored this approach further, both theoretically and computationally. However encouraging, these advances do not vanquish algebraic difficulties in using high-order exponential family models for advanced statistical analyses of directional and axial data.

This paper takes a fully nonparametric approach which does not force upon data the restricted distributional shapes expressed by low-order exponential family models. The nonparametric methods are relatively simple to formulate algebraically. Their sampling distribution theory relies on nonparametric bootstrap asymptotics and their implementation is through Monte Carlo resampling approximations.

\section{Estimators and Confidence Sets} \label{sec2} 
We consider first nonparametric estimators and confidence sets for a 
directional distribution and its mean direction and dispersion. Modifications
to handle axial distributions and functionals are treated subsequently.

\subsection{Directional Data} \label{subsec2.1}
 Let $X_n=(x_1, x_2, \ldots x_n)$ be a sample of independent 
identically distributed random directions each having distribution $P$ on $S_q$. The distribution of $X_n$ is designated as $P^n$. Sought are 
confidence sets for $T(P)$, where $T$ is a functional defined 
on $P$ such as the directional mean or dispersion defined in the Introduction or the directional distribution $P$ itself.  Let $\mathcal{T}$ denote the range of $T(P)$. A confidence set $C$ is a random subset of $\mathcal{T}$ based on the sample $X_n$ such that the \textit{coverage probability} $P^n (C \ni T(P)) = \beta$ for a specified level $\beta$.

The approach in this paper is to derive confidence sets from a real-valued \textit{root} $R_n(X_n,T(P))$.  A classical pivot, such as a t-statistic in normal model theory, is a root whose distribution does not vary with the unknown distribution of the data. However desirable, pivots are rare in statistics. Bootstrap theory enables working with a large class of roots as if they were approximate pivots. Let  $H_n(x,P) = P^n(R_n(X_n, T(P)) < x)$ be the left-continuous cdf of the root $R_n(X_n,T(P))$. A root-based confidence set for $T(P)$ is  then $C= \{t \in \mathcal{T} \colon R_n(X_n,t) \le c(\beta)\}$. In the oracle world where $P$ is known and cdf $H_n(x,P)$ is continuous in $x$, the ideal choice of critical value is $c(\beta) = H_n^{-1}(\beta,P)$, the largest $\beta$-th quantile of the cdf. By construction, this oracular confidence set $C$ has coverage probability $\beta$.

The foregoing motivates a nonparametric bootstrap confidence set for $T(P)$. Let $\Phat_n$ be the empirical distribution of the sample $X_n$. Estimate the left-continuous cdf $H_n(\cdot,P)$ by the theoretical \textit{bootstrap} cdf $\Hhat_B(\cdot) = H_n(\cdot, \Phat_n)$. 
Let $\Hhat_B^{-1}(\beta)$ be its largest $\beta$-th quantile. 
The corresponding \textit{bootstrap confidence set} is then
\begin{equation}
C_B = \{t \in \mathcal{T} \colon R_n(X_n,t) \le \Hhat_B^{-1}(\beta)\}
       = \{t \in \mathcal{T} \colon \Hhat_B(R_n(X_n,t) ) \le \beta)\}.
\label{2.1}
\end{equation}

The theoretical bootstrap distribution $\Hhat_B$ is a random probability 
distribution which can be reinterpreted as a conditional distribution. 
Consider the \textit{bootstrap world} in which the true directional model is 
$\Phat_n$, the functional of interest is $T(\Phat_n)$, and the sample $X_n^*$ is drawn from $\Phat_n$. More precisely, the conditional distribution of $X_n^*$ given $X_n$ is the joint distribution $\Phat_n^n$. Then $\Hhat_B$ is the conditional cdf of $R_n(X_n^*, T(\Phat_n))$ given $X_n$. This view of the matter motivates Monte Carlo sampling in the bootstrap world to approximate the theoretical bootstrap distribution $\Hhat_B$.  An approximation to the bootstrap confidence set $C_B$ replaces $\Hhat_B^{-1}(\beta)$ in \eqref{2.1} with a suitable quantile of the approximate bootstrap distribution. See \cite{bib20} for an important refinement. 

The following is a template for checking when the coverage probability  $P^n(C_B \ni T(P))$ converges to the design level $\beta$ as $n \rightarrow \infty$.  It is a useful subcase of a more general theorem in \cite{bib2}.
\begin{theorem} 
\label{thm1}
Let $\rho$ be a metric for convergence of distributions on $R^q$. 
Suppose that for every distribution $P$ on $S_q$:
\begin{enumerate}
\item[(a)] $\{\Phat_n\}$ is a sequence of estimators such that $\rho(\Phat_n,P) \rightarrow 0$ in $P^n$ probability as $n \rightarrow \infty$;
\item[(b)] (Triangular array convergence). For any sequence of distributions $\{P_n\}$ such that $\rho(P_n,P) \rightarrow 0$ as $n \rightarrow\infty$, the cdfs $\{H_n(\cdot,P_n)\}$ converge weakly to $H(\cdot,P)$;
\item[(c)] The limit cdf $H(x,P)$ is continuous in x.
\end{enumerate}
\noindent
Then 
$\sup_x|\Hhat_B(x) - H(x,P)| \rightarrow 0$ in $P^n$ probability as $n \rightarrow \infty$ and, for $\beta$ in $(0,1)$,
\begin{equation}
\lim_{n \rightarrow\infty} P^n(C_B \ni T(P)) = \beta.
\end{equation}
\end{theorem}

\noindent\textbf{Proof} Conditions (b), (c), and P\'olya's theorem on weak convergence to a continuous cdf imply that
\begin{equation}
\sup_x|H_n(x,P_n) - H(x,P)| \rightarrow 0
\end{equation}
for any sequence of distributions $\{P_n\}$ such that $\rho(P_n,P) \rightarrow 0$ as $n \rightarrow\infty$. In view of condition (a) and the definition of the bootstrap cdf $\Hhat_B(x)$,
\begin{equation}
\sup_x|\Hhat_B(x) - H(x,P)| \rightarrow 0 \mathrm{\ in\ }P^n \mathrm{\ probability\ as\ }n \rightarrow\infty.
\end{equation}
Consequently,
\begin{equation}
|\Hhat_B(R_n(X_n,T(P)) - H(R_n(X_n,T(P)),P)| \rightarrow 0 \mathrm{\ in\ }P^n \mathrm{\ probability\ as\ }n \rightarrow\infty.
\label{2.01}
\end{equation}
By condition (b), as $n \rightarrow\infty$, the roots $R_n(X_n,T(P))$ converge in distribution under $P^n$ to a random variable $R$ with cdf $H(\cdot,P)$.  From the cdf continuity condition (c), the transformed roots $H(R_n(X_n,T(P)),P)$ converge in distribution to $H(R,P)$, which has a Uniform(0,1) distribution. Thus, from \eqref{2.01}, the distribution of 
$\Hhat_B(R_n(X_n,T(P)))$ converges weakly to a Uniform(0,1) distribution. From this and \eqref{2.1},
\begin{equation}
\lim_{n \rightarrow\infty} P^n(C_B \ni T(P)) 
=\lim_{n \rightarrow\infty}P(\Hhat_B(R_n(X_n,T(P))) \le \beta) = \beta
\end{equation}
as assserted in the Theorem.

We now construct nonparametric estimators and bootstrap confidence sets for the mean direction $\mu(P)$, the directional dispersion  $\delta(P)$, and the directional distribution $P$ itself. As needed it is assumed that 
$|m(P)| > 0$. Condition (a) in Theorem~\ref{thm1} is satisfied by the empirical distribution $\Phat_n$ when $\rho$ metrizes weak convergence. Using \eqref{1.2} and \eqref{1.3}, nonparametric estimators of $\mu(P)$ and $\delta(P)$ are
\begin{equation} 
\muhat_n=\mu(\Phat_n) = m(\Phat_n)/|m(\Phat_n)|
= (\Sigma_{i=1}^n x_i)/|\Sigma_{i=1}^n x_i|
\label{2.2}
\end{equation}
and
\begin{equation} 
\deltahat_n = \delta(\Phat_n) = 2(1-|m(\Phat_n)|) 
= 2(1-|n^{-1}\Sigma_{i=1}^n x_i|).
\label{2.3}
\end{equation}
Both estimators converge in probability to their respective targets as $n$ 
tends to infinity.

\subsubsection{Confidence cone for mean direction} \label{subsubsec2.1.1}
A root for this construction is
\begin{equation} 
R_n(X_n,\mu(P)) = n|\muhat_n - \mu(P)|^2 = 2n(1 - \muhat_n'\mu(P)).
\label{2.4} 
\end{equation}
The bootstrap confidence set for  $\mu(P)$ is
\begin{equation}
C_B = \{\mu \in S_q\colon R_n(X_n,\mu) \le \Hhat_B^{-1}(\beta)\}.
\label{2.5}
\end{equation}
Geometrically this confidence set is a circular cone on $S_q$ about the 
estimated mean direction $\muhat_n$. Because the root \eqref{2.4} meets the conditions of Theorem~\ref{thm1} when $\rho$ metrizes weak convergence, the confidence cone has asymptotic coverage probability $\beta$. To approximate the cone numerically through Monte Carlo resampling of $\Phat_n$, it suffices to bootstrap the inner product $\muhat_n'\mu(P)$. A detailed theoretical and computational study of this confidence cone is \cite{bib5}. Three-dimensional directional data, the estimated mean direction, and the confidence cone for mean direction are usually plotted in an area-preserving Lambert projection. See \cite{bib8} for a brief derivation and discussion of this projection.

\subsubsection{Confidence interval for directional dispersion}
\label{subsubsec2.1.2}
A root for a two-sided confidence interval is
\begin{equation}
R_n(X_n,\delta(P)) = n^{1/2}|\deltahat_n - \delta(P)| 
= 2n^{1/2}||m(\Phat_n)| - |m(P)||.
\label{2.6}
\end{equation}
The bootstrap confidence interval for $\delta(P)$ is
\begin{equation}
C_B = \{\delta \in [0,2] \colon R_n(X_n,\delta) \le \Hhat_B^{-1}(\beta)\}.
\label{2.7}
\end{equation}
Because its root fulfills the conditions of Theorem \ref{thm1}, the confidence 
interval has asymptotic coverage probability $\beta$. Any one-to-one mapping of $\delta(P)$ and $\deltahat_n$ can be used for an alternative root.

\subsubsection{Confidence set for a directional distribution}
\label{subsubsec2.1.3}
The task is to construct a confidence set for the directional distribution $P$ on $S_q$ that is unaffected by orthogonal transformations of the Cartesian coordinate system in which the directions $X_n$ are recorded. For every $(s,t)\in S_q \times R^1$, let $A(s,t)$ be the half-space in $R^q$
\begin{equation}
A(s,t)=\{z\in R^q \colon s'z \le t\}.
\label{2.8}
\end{equation}
The half-spaces separate probabilities on $R^q$ in the sense that if $P_1(A)$ coincides with $P_2(A)$ for all half-spaces $A$, then $P_1$ and $ P_2$ agree on all Borel sets.

The half-space metric between distributions $P_1$ and $P_2$ on $R^q$ is defined in \cite{bib12} as
\begin{equation}
h(P_1,P_2) = \sup\{|P_1(A(s,t)) - P_2(A(s,t))| \colon (s,t) 
\in S_q \times R^1\}.
\label{2.9}
\end{equation}
Let the confidence set root for a distribution $P$ on $S_q$ be
\begin{equation}
R_n(X_n,P) = n^{1/2}h(\Phat_n, P).
\label{2.10}
\end{equation}
The corresponding bootstrap confidence set for $P$ is
\begin{equation}
C_B = \{P \mathrm{\ on\ } S_q \colon R_n(X_n,P) \le \Hhat_B^{-1}(\beta)\}.
\label{2.11}
\end{equation}
The root \eqref{2.10} satisfies the conditions of Theorem~\ref{thm1} by the bootstrap arguments made in \cite{bib4}. Consequently, this confidence set has asymptotic coverage probability $\beta$. For the sphere $S_3$, $C_B$ consists of simultaneous confidence intervals for the probabilities assigned by $P$ to every cap on the sphere. For the circle $S_2$, caps are replaced by arcs.

Stochastic approximation of the half-space distance $\rho(P_1, P_2)$ for the bootstrap resampling may be accomplished as follows. Let $s_1,s_2, \ldots s_{k_n}$ be independent identically distributed random unit vectors which are uniformly distributed on $S_q$ and independent of $X_n$. Define the approximation through
\begin{equation}
h_n(P_1,P_2) = \max_{1 \le k\le k_n} \sup \{|P_1(A(s_k,t)) - P_2(A(s_k,t))| \colon t \in R^1\}.
\label{2.12}
\end{equation}
If $\lim_{n \rightarrow\infty}k_n = \infty$, then $\lim_{n\rightarrow\infty}
h_n(P_1,P_2) = h(P_1,P_2)$ w.p.1 and the rate of convergence in probability is exponential. This result motivates Monte Carlo approximations to the half-space distance $h(P_1,P_2)$. 

Section 3.2 of \cite{bib4} presents a case study on geological data in $S_3$ of Monte Carlo approximation to the half-space confidence set \eqref{2.11}. It is found that the Fisher distribution fitted to the data by maximum likelihood fails to lie within bootstrap confidence set $C_B$ at nominal level .985.  

\subsection{Axial Data} \label{subsec2.2}
Let $Y_n = \{x_1x_1', x_2x_2', \ldots x_n x_n'\}$ be a sample of independent identically distributed random axes that are expressed as rank 1 projections, each having distribution $Q$. The distribution of $Y_n$ is designated as $Q^n$. Of interest are confidence sets for a functional $T(Q)$ such as the mean axis or axial dispersion defined in the Introduction or the axial distribution $Q$ itself. The constructions parallel those given for directional data in subsection \ref{subsec2.1}.

It is assumed that the largest eigenvalue of the matrix $M(Q)$ defined in equation \eqref{1.4} is unique.  The empirical distribution $\Qhat_n$ of $Y_n$ estimates $Q$. Using equation \eqref{1.5}, the nonparametric estimator $\pm\epsilonhat_n$ of the mean axis $\pm\epsilon(Q)$ is the sign ambiguous eigenvector of
\begin{equation} 
\Mhat_n = M(\Qhat_n) = n^{-1}\Sigma_{i=1}^n x_ix_i'.
\label{2.13}
\end{equation}
that is associated with the largest eigenvalue $\lambdahat_{\max,n}$ of $\Mhat_n$. From equation \eqref{1.6}, the estimator of the axial dispersion 
$\gamma(Q)$ is then
\begin{equation} 
\gammahat_n = \gamma(\Qhat_n) = 2(1- \lambdahat_{\max,n}).
\label{2.14}
\end{equation}

\subsubsection{Confidence double cone for mean axis} \label{subsubsec2.2.1}
In the notation of the preceding paragraph, a root for this construction is
\begin{equation}
 R_n(Y_n,\epsilon(Q)) =  n||\epsilonhat_n\epsilonhat_n'-\epsilon(Q)\epsilon(Q)'||^2 
 = 2n(1 - (\epsilonhat_n'\epsilon(Q))^2).
 \label{2.15}
\end{equation}
The bootstrap confidence set for  $\mu(P)$ is
\begin{equation}
C_B = \{\pm\epsilon \colon\epsilon\in S_q, R_n(Y_n,\epsilon) \le \Hhat_B^{-1}(\beta)\}.
\label{2.16}
\end{equation}
Geometrically this confidence set is a circular double cone on $S_q$ about the estimated mean axis $\pm\epsilonhat_n$. That the conditions of Theorem \ref{thm1} with $P$ replaced by $Q$ are satisfied is established by  nonparametric bootstrap arguments presented in \cite{bib3} for the eigenprojections of a sample covariance matrix. To approximate the double cone numerically through Monte Carlo resampling from $\Qhat_n$, it suffices to bootstrap the squared inner product $(\epsilonhat_n'\epsilon(Q))^2$.

\subsubsection{Confidence interval for axial dispersion} \label{subsubsec2.2.2}
Using \eqref{2.14}, a root in this case is
\begin{equation}
R_n(Y_n, \gamma(Q)) = n^{1/2}|\gammahat_n - \gamma(Q)|
= 2n^{-1/2}|\lambdahat_{\max,n} - \lambda_{\max}(Q)|.
\label{2.17}
\end{equation}
The bootstrap confidence interval for $\gamma(Q)$ is
\begin{equation}
C_B = \{\gamma \in [0,2] \colon R_n(X_n,\gamma) \le \Hhat_B^{-1}(\beta)\}.
\label{2.18}
\end{equation}
By the nonparametric bootstrap arguments in \cite{bib3}, \cite{bib21} for the eigenvalues of a sample covariance matrix, the root fulfills the conditions of the axial analog to Theorem \ref{thm1} and the confidence interval has asymptotic coverage probability $\beta$. 

\subsubsection{Confidence set for an axial distribution} \label{subsubsec2.2.3}
Let $S_q^+$ denote the hemisphere that consists of all unit vectors in $R^q$ whose first nonzero component is positive. A random direction $x$ with distribution $P$ on the hemisphere $S_q^+$ identifies a random axis $\pm x$ and its distribution. A sample $X_n = (x_1,x_2, \ldots, x_n)$ of independent identically random directions in $S_q^+$, each having distribution $P$, identifies a sample of random axes. 

Let $\Phat_n$ denote the empirical distribution of $X_n$. A bootstrap confidence set for $P$ with asymptotic coverage probability $\beta$ is constructed as in section \ref{subsubsec2.1.3} and is supported by the results stated there. It suffices to recognize that $P$ and $\Phat_n$ are supported on the hemisphere $S_q^+$ rather than on the full sphere. This nonparametric confidence set for an axial distribution can be used to assess the trustworthiness of Bingham distribution fits to axial data.

\section{Further Developments} \label{sec3}
Suppose more generally that $P$ is a distribution supported on a specified compact subset $\K \in R^q$. Let $x$ have distribution $P$ and let $m(P) = \E(x)$ be its expectation. Note that for any $\mu \in \K$,
\begin{equation}
\E|x -\mu|^2 = \E|x-m(P)|^2 + |m(P) - \mu|^2. 
\label{2.19}
\end{equation}
By analogy with \eqref{1.2}, the \textit{mean of $P$ on $\K$} is defined as
\begin{equation}
\mu(P) = \argmin_{\mu \in \K}\E|x - \mu|^2 = \argmin_{\mu \in \K}|m(P) - \mu|^2.
\label{2.20}
\end{equation}
Thus, $\mu(P)$ is the projection in Euclidean norm of $m(P)$ onto 
$\K$. This extends the concepts of mean direction or mean axis treated earlier. By analogy with \eqref{1.3}, the \textit{dispersion of $P$ on $\K$} is defined to be
\begin{equation}
\delta(P) = \min_{\mu \in \K}\E|x- \mu|^2 = \E|x-\mu(P)|^2 = \sigma^2(P) + |m(P) - \mu(P)|^2,
\label{2.21}
\end{equation}
where $\sigma^2(P) = \E|x-m(P)|^2$.

Let $X_n$ be a sample of independent identically distributed random vectors, each having distribution $P$ on $\K$ and let $\Phat_n$ denote their empirical distribution. Then $\muhat_n = \mu(\Phat_n)$ and $\deltahat_n = \delta(\Phat_n)$ estimate the functionals \eqref{2.20} and \eqref{2.21} while $\Phat_n$ estimates $P$.
The half-space confidence sets for $P$ defined in section \ref{subsubsec2.1.3} remain valid in the present setting because the underlying asymptotic theory in \cite{bib4} holds for all distributions on $R^q$. Confidence sets for $\mu(P)$ and $\delta(P)$ on $K$  can be proposed by analogy with those in sections \ref{subsubsec2.1.1} and \ref{subsubsec2.1.2}. Justifying these requires developing arguments akin to Theorem \ref{thm1}. Asymptotic bootstrap theory for extrinsic means on Riemannian manifolds is provided in \cite{bib9} and \cite{bib10} and is extended to further applications in \cite{bib11}.

Estimation of many directional means raises new considerations. A simple directional trend estimator that respects the geometry of $S_q$ is to compute a running average over the time-ordered observed directions, then normalize each of the average vectors to unit length. A more general class of directional trend estimators treated in \cite{bib14} similarly projects onto $S_q$ a symmetric linear transform of the matrix of observed directions. Within this richer class lie extrinsic penalized least squares estimators for multiple mean directions. The supporting asymptotic theory applies affine shrinkage to reduce estimation risk.

To compare two directions $d_1$ and $d_2$ in $S_3$, consider their \textit{relation}: the plane determined by $d_1$ and $d_2$ together with the rotation angle in that plane which takes $d_1$ into $d_2$. Paper \cite{bib16} expresses this pairwise relation in $S_3$ as a unit vector in $S_4$ Thereby it develops nonparametric bootstrap confidence sets for the common plane holding two unknown mean directions and for the rotation angle taking one into the other.  Analogous bootstrap confidence sets hold for pairwise comparison of two mean directions in $2$ dimensions and for pairwise comparison of two mean axes in either $2$ or $3$ dimensions.  Balanced simultaneous confidence sets for all pairwise comparisons among $k > 2$ mean directions or axes are constructed by using the bootstrap generalization in \cite{bib17} of the classical Tukey and Scheff{\'e} simultaneous confidence intervals.

A collaboration among linear algebra, convex optimization, probability theory, bootstrap methods and computer algorithms underlies these developments.

\bibliography{beran-bib}
\end{document}